\documentclass[twocolumn,showpacs,preprintnumbers,amsmath,amssymb]{revtex4}
\usepackage{amsmath}
\usepackage{graphicx}
\usepackage{epsf}
\usepackage{psfrag}
\begin{document}
\newcommand{\be}{\begin{equation}}
\newcommand{\ee}{\end{equation}}
\newcommand{\bq}{\begin{eqnarray}}
\newcommand{\eq}{\end{eqnarray}}

\title{{\bf{Dispersion and uncertainty in multislit matter wave diffraction}}}
\date{\today}

\author{G. Glionna} \email []{gglionna@fisica.ufmg.br}
\author{A. H. Blin} \email[]{alex@teor.fis.uc.pt}
\author{B. Hiller}\email[]{brigitte@teor.fis.uc.pt}
\author{M. C. Nemes}\email[]{carolina@fisica.ufmg.br}
\author{Marcos Sampaio}\email[]{sampaio@th.u-psud.fr}
\author{A. F. R. de Toledo Piza}\email[]{piza@fma.if.usp.br}
\affiliation{Federal University of Minas Gerais -
Physics Department - ICEx \\ P.O. BOX 702, 30 161-970, Belo Horizonte MG - Brazil}
\affiliation{Physics Department, CFT, University of Coimbra, 3000, Portugal}
\affiliation{Institute of Physics, University of S\~ao Paulo, P. O. Box 66 318,\\
5315-970, S\~ao Paulo, Brazil}
\affiliation{Universit\'e Paris Sud XI - LPT - Centre d'Orsay, 91 405, France.}

\begin{abstract}

We show that single and multi-slit experiments involving matter waves
may be constructed to assess correlations between the position and
momentum of a single free particle. These correlations give rise to
position dependent phases which develop dynamically and may play an
important role in the interference patterns. For large enough
transverse coherence length such interference patterns are noticeably
different from those of a classical dispersion free wave.

\end{abstract}


\maketitle

Fundamental aspects of quantum mechanics are revealed by diffraction
experiments with particles. Much work has been devoted to the matter
considering electron \cite{1}, neutron \cite{2} and more recently
large molecule \cite{3,4} diffraction from multi-slit gratings. In the
present contribution we are concerned with a double diffraction
problem. The question to be addressed is how sensitive are the
measured interference patterns to the dispersive dynamics of particle
propagation {\it before} it reaches the grating and proceeds from
there to the screen. In order to avoid having to consider the
production mechanism explicitly (see ref.~\cite{5,6} for that
purpose), we assume that there is a first slit after the beam is
collimated  which determines its transverse correlation
length. Also, as done in ref.~\cite{7}, we assume that the
longitudinal (beam direction) wave packet localisation is sharp enough
compared with the flight path so that the time-of-flight approximation
is valid. In other words, the flight paths involved in the problem are
much larger than the position spread in longitudinal
direction. Coherence loss mechanisms are neglected, since their
effects are well known. In this idealised scenario a Gedanken
experiment is described in terms of the time evolution of an
initially Gaussian wave-packet which travels freely from the first
slit to the multi-slit grating.

This model sheds light on specific matter wave dispersion effects,
notably the dynamical evolution of the position-momentum
correlations. For a real initial Gaussian wave-packet, these
correlations appear in the form of a {\it position dependent} phase as
soon as the particle leaves the first slit. The importance of this
position dependent phase depends on the ratio between the time of
flight and an intrinsic time $\tau=m\sigma_0^2/\hbar$, $m$ being the
particle mass and $\sigma_0$ is the initial width of the
wave-packet. Therefore if the particle travels long enough before it
reaches the multi-slit grating, it will arrive at each slit with a
different phase. As will be shown in what follows, this may radically
alter the interference patterns observed on the screen.

In order to review the essential dispersive dynamical effects we
discuss the time evolution in the transverse direction, from the first
slit at $t = 0$ to the grating, of a Gaussian wave packet given by

\be
\varphi (x,0) = \Bigg( \frac{1}{\sigma_0 \sqrt{\pi}}
\Bigg)^{1/2} \exp \Bigg({\frac{-x^2}{2 \sigma_0^2}}\Bigg) \,,
\ee

\noindent where $\sigma_0$ is the width of the first slit. Its time
evolution according to Schroedinger's equation will yield for the wave
packet just before the grating

\begin{eqnarray}
\varphi(x, T) &=& \Bigg( \frac{1}{B(T) \sqrt{\pi}} \Bigg)^{1/2}
\exp \Bigg[ - \frac{x^2}{2 B^2(T)}
\Bigg( 1 - \frac{i \hbar T}{m \sigma_0^2} \Bigg)\Bigg] \,\equiv
\nonumber \\ &\equiv& \widetilde{\varphi} (x,T) \exp(i S (x,T))\,,
\nonumber
\label{eq:wf}
\end{eqnarray}

\noindent where

\bq
\widetilde{\varphi} (x, T) &=& \left(\frac{1}{B(T) \sqrt{\pi}}
\right)^{1/2} \exp \left[ - \frac{x^2}{2 B^2(T)} \right]
\,,\nonumber \\
B^2(T) &=& \sigma_0^2 \Big( 1 + \frac{\hbar^2 T^2}{m^2 \sigma_0^4}
\Big)
\eq

\noindent and

\be
S = \frac{x^2}{2 B^2(T)} \frac{\hbar T}{m \sigma_0^2} \equiv
\frac{x^2}{2 B^2(T)} \frac{T}{\tau_0} \,.
\ee

\noindent Notice that the position dependent phase $S$ contains the
time scale $\tau_0=\frac{m \sigma_0^2}{\hbar}$. The ratio $T/\tau_0$
will determine the importance of this $x-$dependent phase to the
interference pattern. In the experimental setups using fullerene
molecules \cite{3} $T/\tau_0 \approx 10^{4}$ which is also the
condition for Frauhoffer diffraction (see ref.~\cite{7}).  The time
scale $\tau_0$ is fundamentally determined by Heisenberg's uncertainty
relation, given the initial position dispersion $\Delta x (0) =
\sigma_0/\sqrt{2}$. In fact, the corresponding momentum dispersion is
$\Delta p = \hbar/(\sigma_0 \sqrt{2})$. Because the momentum is a
constant of motion this momentum spread will be preserved in
time. Both $\Delta x$ and $\Delta p$ constitute {\it{intrinsic}}
properties of the initial wave packet, in terms of which the time
scale $\tau_0$ is expressed as

\be
\tau_0 = \frac{\Delta x (0)}{(\Delta p)/m}\,.
\ee

\noindent The numerator in the above relation represents the spatial
dimensions of the initial wave packet, whilst the denominator stands
for the scale of {\it{velocity differences}} enforced by the
uncertainty principle. Therefore the time scale $\tau_0$ corresponds
essentially to the time during which a distance of the order of the
wave packet extension is traversed with a speed corresponding to the
dispersion in velocity. It can therefore be viewed as a characteristic
time for the ``aging'' of the initial state, which consists in
components with larger velocities (relatively to the group velocity of
the wave packet) concentrating at the frontal region of the packet.
This can be seen explicitly by deriving the velocity field associated
with the phase $S$ in equation \ref{eq:wf}, which reads

\be
v (x, T) = \frac{\hbar}{m} \frac{\partial S}{\partial x} =
\frac{T x}{\tau_0^2 + T^2}\,.
\ee

\noindent This expression shows that for $T>0$ the initial velocity
field $v(x, 0)=0$ varies linearly with respect to the distance from center of the
wave packet ($x = 0$).

Next we relate quantitatively this ``ageing'' effect to
position-momentum correlations. This is readily achieved  using the
generalised uncertainty relation devised by Schroedinger \cite{9},
which is expressed in this case  in terms of the determinant of the
covariance matrix $\Sigma$

\[
{\rm det}\Sigma\equiv{\rm det}\left(\begin{array}{cc}
\Delta x^2 & \frac{1}{2}\langle\hat{x}\hat{p}+\hat{p}\hat{x}\rangle
\\ \frac{1}{2}\langle\hat{x}\hat{p}+\hat{p}\hat{x}\rangle & 
\Delta p^2 \end{array} \right)\geq\frac{\hbar^2}{4}\,.
\]

\noindent For the minimum uncertainty wave packet of equation
\ref{eq:wf} we obtain, {\it at all times},

\be
{\rm det} \left(
\begin{array}{cc}
\frac{B^2(T)}{2} & \frac{ \hbar T}{2 \tau_0} \\
\frac{\hbar T}{2 \tau_0} & \frac{\hbar^2}{2 \sigma_0^2}\\
\end{array} \right) = \frac{\hbar^2}{4} \, .
\ee

\noindent The Gaussian wave packet therefore saturates Schroedinger's
uncertainty relation at all times. We can thus describe this result by
saying that the time dependence of the {\it Heisenberg} uncertainty
relation (obtained by dropping the off-diagonal elements of the
covariance matrix) which is due, in this case, to the dispersive
increase of the wave packet width in time, just reflects the $x-p$
correlation process. The relevant quantity in this connection is the
correlation matrix element

\be
\frac{\langle \hat{x} \hat{p} + \hat{p}\hat{x} \rangle}{2} =
-\hbar \frac{T}{2 \tau_0}\,.
\ee

Recently an interesting experiment \cite{8} has been performed in
order to study the Heisenberg uncertainty relation for fullerene
molecules using the fullerene $C_{70}$ and measuring the momentum
spread after the passage through a narrow slit with a variable width
(down to $70 \, nm$). The results are interpreted in the light of
Heisenberg's uncertainty relation. The results are sumarized by the
empirical relation

\begin{equation}
\Delta p=\frac{C h}{\Delta x}\,,
\label{eq:ir}
\end{equation}

\noindent where $C= 0.89$ and $h = 2 \pi \hbar$. Having analytical
expressions for each of the elements of the covariance matrix $\Sigma$
and using the empirical value $\Delta x \Delta p=0.89h$ we can
evaluate $\langle x p + p x \rangle $. Using the values of the
experiment of ref. \cite{8} we get $\langle x p + p x
\rangle\approx 11.14\hbar$.

\vspace{.5cm}

\begin{figure}
\begin{center}
\includegraphics[scale=0.5]{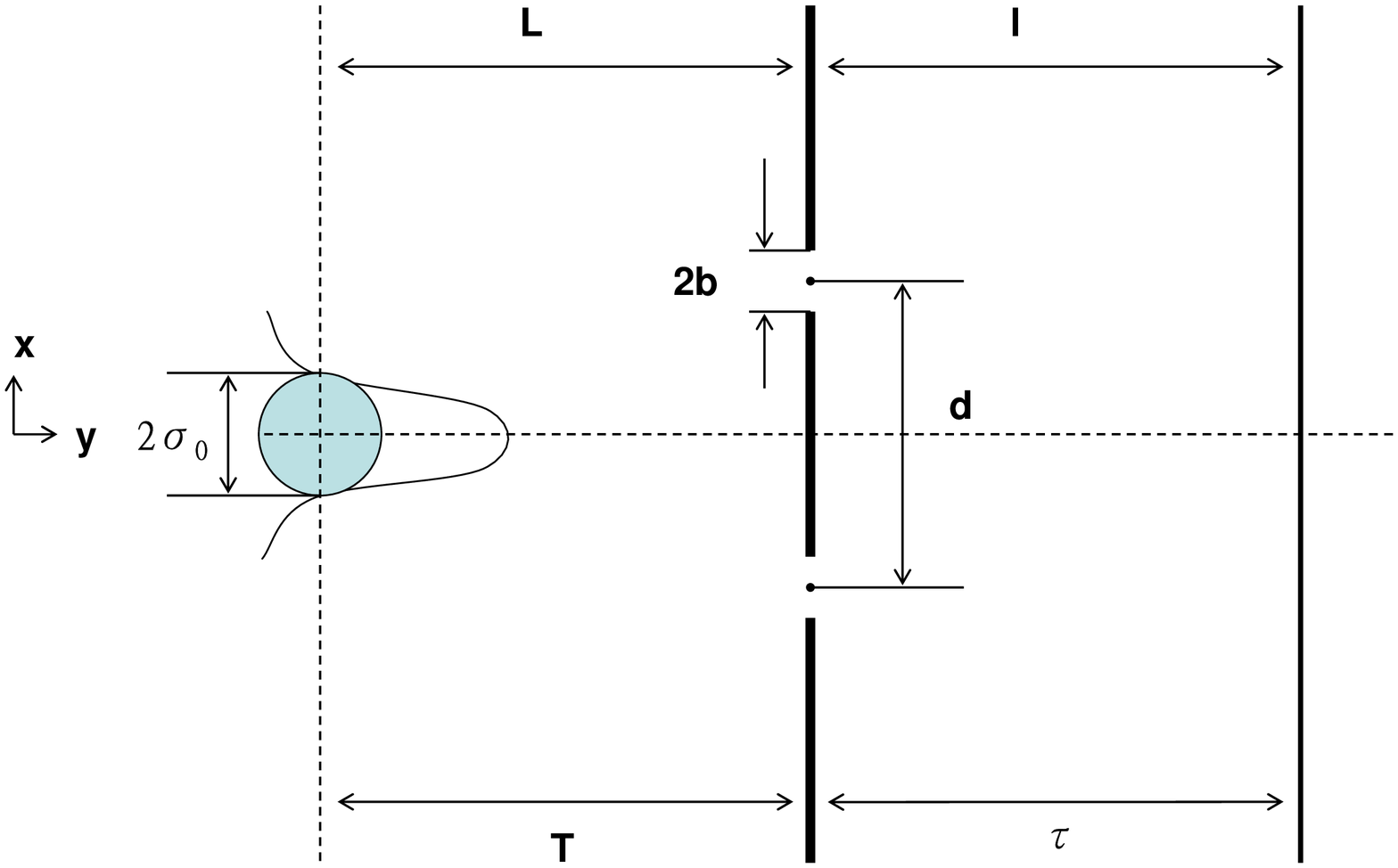}
\caption{Sketch of a double diffraction arrangement for a two-slit
grating. Parameters: $\sigma_0=0.5\times 10^{-5}$m, $d=10^{-7}$m,
$b=1.8\times 10^{-8}$m, $L=0.1$m, $l=1.25$m, and $T=L/v$, $\tau=l/v$,
where $v$ is the velocity of the molecules.}
\label{fig3}
\end{center}
\end{figure}

\textbf{Two slit grating.} We next consider the double diffraction
experiment for a two slit grating following the first slit (see
fig. \ref{fig3}). In this case the intensity at the screen is given by
\cite{7}

\be
I(x) = \bigl\lvert
\Psi_{+}(x,T,\tau)+\Psi_{-}(x,T,\tau)\bigr\rvert^2\ .
\ee

\noindent where

\bq
\Psi_{\pm} (x,T,\tau) &=& \int_{-\infty}^{+\infty}
dx_i\int_{-\infty}^{+\infty} dw \, K(x,T+\tau;w,T) \times \nonumber \\
&\times& G(w \pm d/2)
K(w,T;x_i,0) \varphi (x_i)
\eq

\noindent and

\begin{align}
K(z,t;w,t_0) & = \sqrt{\frac{m}{2 \pi i\hbar(t-t_0)}}\exp
\left[i\frac{m (z-w)^2}{2\hbar(t-t_0)}\right]\ ,\\
G(w) & =  \exp\left[-\frac{w^2}{2 b^2}\right]\,,\ \mbox{and}
\label{eq:G} \\ \varphi (x_i) & = \frac{1}
{\sqrt{\sigma_0\sqrt{\pi}}} \exp\left[-\frac{x_i^2}
{2\sigma_0^2}\right].
\label{eq:iwf}
\end{align}

\noindent The kernel $K(z, t, w, t_0)$ is the free propagator for the
particle, the functions $G(w\pm d/2)$ describe the double slit
apertures which are taken to be Gaussian of width $b$ separated by a
distance $d$; the width of the first slit is $\sigma_0$, $m$ is the
mass of the particle, $T$ ($\tau$) is the time of flight from the
first slit to the double slit. Parameter values are taken from
ref. \cite{8}.

\begin{figure}
\begin{center}
\psfrag{p}[cc]{Position, $x\ ({\rm \mu m})$}
\psfrag{i}[cc]{Intensity}
\psfrag{s}[l][l][1]{$\sigma_0\ ({\rm \mu m})$}
\psfrag{l1}[l][l][0.8]{6.0}
\psfrag{l2}[l][l][0.8]{0.02}
\psfrag{l3}[l][l][0.8]{0.0175}
\psfrag{l4}[l][l][0.8]{0.013}
\psfrag{x1}[ct][l][0.8]{-40}
\psfrag{x2}[ct][l][0.8]{-20}
\psfrag{x3}[ct][l][0.8]{0}
\psfrag{x4}[ct][l][0.8]{20}
\psfrag{x5}[ct][l][0.8]{40}
\psfrag{i1}[rc][l][0.8]{0.2}
\psfrag{i2}[rc][l][0.8]{0.4}
\psfrag{i3}[rc][l][0.8]{0.6}
\psfrag{i4}[rc][l][0.8]{0.8}
\psfrag{i5}[rc][l][0.8]{1.0}
\includegraphics[scale=0.6,angle=0]{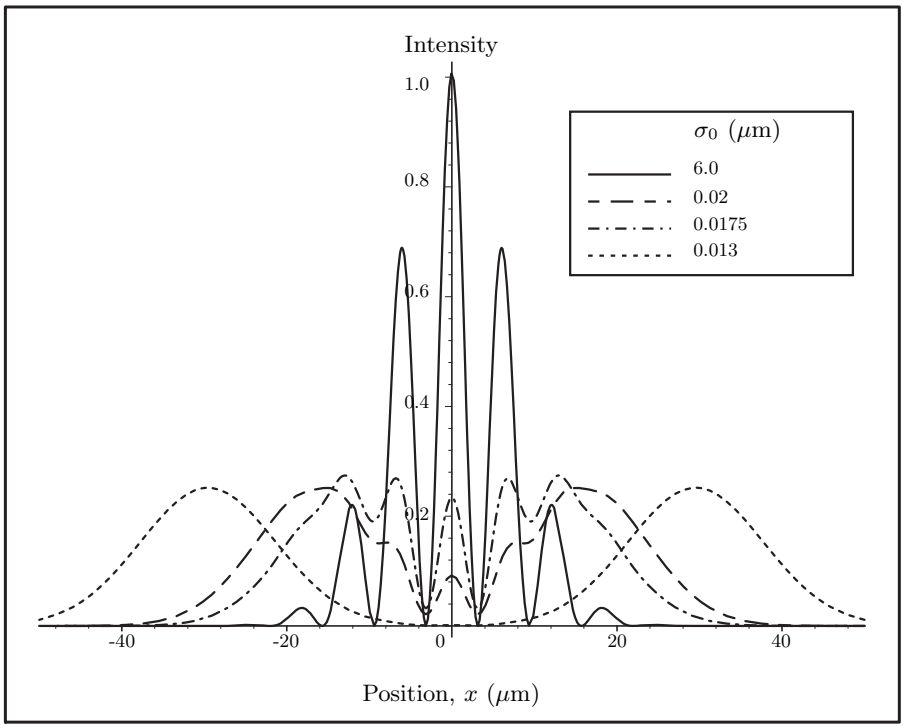}
\caption{Diffraction patterns for the arrangement of fig. \ref{fig3},
for different values of the width of the first slit $\sigma_0$. As the intensity at $x=0$ diminishes we have $\sigma_0 = 6.0 \, \mu m$, $\sigma_0 = 0.02 \, \mu m$, $\sigma_0 = 0.0175 \, \mu m$, $\sigma_0 = 0.013 \, \mu m$, respectively. Recall that the distance between two slits is $\approx 0.1 \, \mu m$.}
\label{fig5}
\end{center}
\end{figure}

Let us now allow for wave packet ``ageing'', i.e., for significant
transverse spreading with the accompanying $x-p$ correlation effects
{\it before} reaching the two slit grating. This can be achieved in a
variety of ways, but we choose for simplicity to make $\sigma_0$
smaller while keeping the width just before the slits fixed at $1 \mu
m$. The result is shown in figure \ref{fig5} for different values of
$\sigma_0$, other parameters remaining unchanged. Note the qualitative
changes of the interference pattern as the $x-p$ correlations grow,
implying increasing phase difference between the contributions of the
impinging wave packet at the two slits with decreasing $\sigma_0$.

\vspace{.5cm}

\textbf{Multi-slit grating.} An alternate way to bring about the
effects of quantum mechanical dispersion with fixed $\sigma_0$ is to
consider diffraction by an increasing number of equally spaced
diffraction slits instead of just two. In order to explore this
strategy we evaluate

\begin{equation}
I(x) = \lvert \sum_{n=0}^{N-1}\Psi_n(x, T, \tau)\rvert^2
\label{intt}
\end{equation}
 where 
 \bq
&&\Psi_n(x, T, \tau) = \int_{-\infty}^{+\infty}dw  \int_{-\infty}^{+\infty}dx_i K(x, T+\tau ; w, T) \nonumber \\  
&& G(w-X_n)K(w, T ; x_i, 0) \varphi (x_i)
 \eq
with $X_n = X_0 - (N -1)\frac{d}{2} + n d $ for $N$ slits centered around $X_0 = 0$.  As discussed before, the wave function at the grating is given by

\begin{equation}
\Psi(x,t) \propto\exp\left[-\frac{x^2}{2B^2(t)}\Bigl(1 - \frac{i\hbar t}
{m\sigma_0^2}\Bigr)\right]
\label{eqn:phase}
\end{equation}

\noindent where $B(t)=\sigma_0\sqrt{1+(\hbar^2
t^2)/(m^2\sigma_0^4)}$. Note again the second term in the exponent
giving rise to the quantum dispersive phase which will be different at
each slit position. An estimate of the number of slits above which the
effects of the phase in equation (\ref{eqn:phase}) become effective
(assuming an infinite transverse correlation length) is

\be
N^2 d^2 \frac{\hbar^2 T^2}{m^2\sigma_0^4}.
\ee

\begin{figure}
\begin{center}
\includegraphics[scale=0.5]{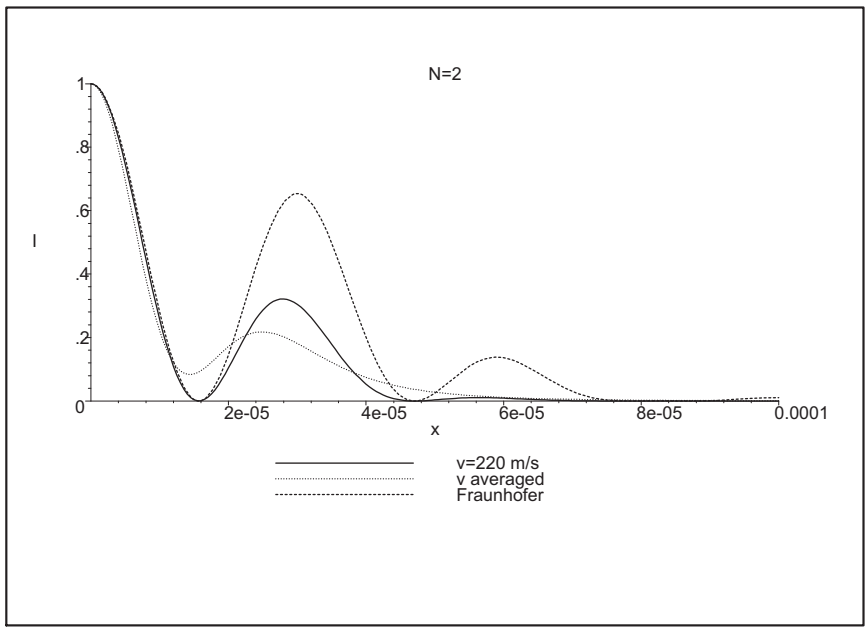}
\caption{Intensity pattern for two slits and parameters of
fig. \ref{fig3}.}
\label{fig6}
\end{center}
\end{figure}

\begin{figure}
\begin{center}
\includegraphics[scale=0.5]{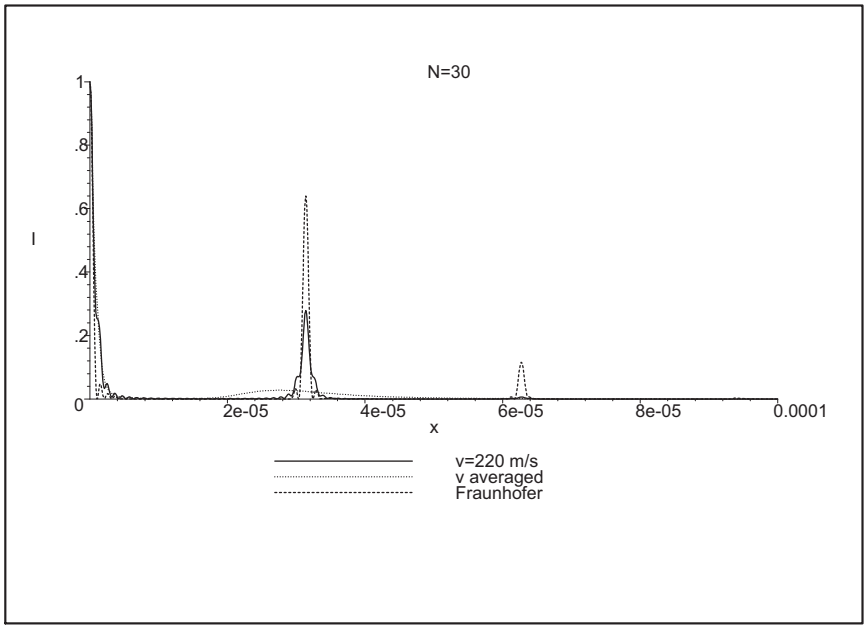}
\caption{Intensity pattern for thirty slits and parameters of
fig. \ref{fig3}.}
\label{fig9}
\end{center}
\end{figure}

\begin{figure}
\begin{center}
\includegraphics[scale=0.5]{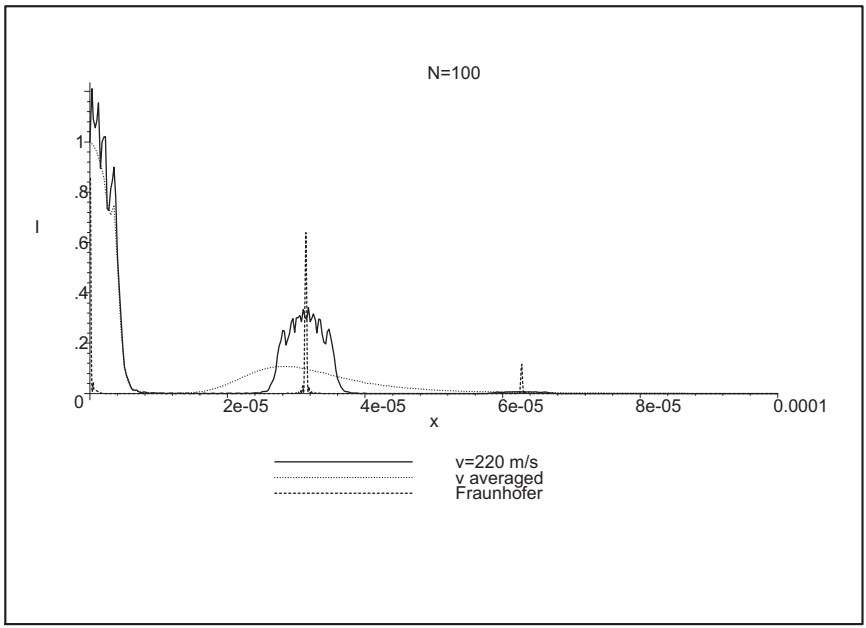}
\caption{Intensity pattern for a hundred slits and parameters of
fig. \ref{fig3}.}
\label{fig10}
\end{center}
\end{figure}

\noindent With the parameter values of ref. \cite{8} and
$\sigma_0=0.5\times 10^{-5}$m one obtains $N\sim 30$. The intensity $I(x)$
is depicted as a function of transverse position $x$ in
figures \ref{fig6} to \ref{fig10} for different values of the number
of slits $N$ (only half of the symmetric interference pattern is
shown). The parameters used correspond to the experimental setup for
$C_{60}$ molecules of reference \cite{3}. In particular we take
$\sigma_0=0.5\times 10^{-5}$m in the initial wave packet
eq.(\ref{eq:iwf}). The grating is characterised by the half width of
the slits $b$ and by the slit spacing $d$, the times $T$ and $\tau$ in
eq.  (\ref{eq:wf}) are calculated from the velocity $v$ of the
molecules and the distance from the first slit to the grating and from
there to the detector position respectively. Each figure shows the
intensity for the most probable velocity $v=200$ m/s (full line) and
an incoherent sum over velocities (dotted line), which takes into
account the experimental spread in the initial velocities of about $60
\%$ as parameterised in reference \cite{3}. The dash-dotted lines
exhibit the corresponding classical Fraunhofer interference pattern
based on the de Broglie wavelength of the $C_{60}$ molecule with
$v=220$ m/s.

The classical Fraunhofer expression for the intensity is very
similar to the one with a definite velocity in the first case, $N = 2$. For the
second case, $N = 30$, deviations from the classical Fraunhofer pattern can be
seen. Note also that with such a large spread in the velocities this
difference will not be experimentally accessible. For larger values
of $N$ as shown in figure \ref{fig10} ($N=100$),  the difference between a
matter wave and a classical wave becomes qualitative.

These are purely quantum effects; in fact they are quantum effects
coming from the first stage of the experiment and the position dependent
phase. Of course these effects could be observed for a smaller number of
slits provided the $x$-$p$ correlations in the first part of the
experiment be large enough.

In summary, for multi-slit diffraction of matter waves we have shown
that in an experiment with large enough transverse correlation length
and small enough incoherence in the initial beam it is possible in
principle to distinguish typical classical wave patterns from typical
quantum matter wave patterns as a function of the number of slits.

\vspace{.5cm}

\textbf{Acknowledgements:} The work of MCN was partially supported by
CNPq, FAPESP, PRAXIS/BCC/4301/94, PRAXIS/FIS/12247/98 and
POCTI/1999/FIS/3530 and MS was supported by CAPES-MEC.

\newpage
.
\newpage

%%%%%%%%%%%%

\end{document}